\documentclass[fleqn,twoside]{article}
\usepackage{espcrc2}
\usepackage{amssymb}

\usepackage{graphicx}
\usepackage{epsfig}
\usepackage{psfrag}

\newcommand{\beq}{\begin{equation}}
\newcommand{\eeq}{\end{equation}}
\newcommand{\beqn}{\begin{eqnarray}}
\newcommand{\eeqn}{\end{eqnarray}}
\newcommand{\bea}[1]{\beq\begin{array}{#1}}
\newcommand{\eea}{\end{array}\eeq}
\newcommand{\eq}[1]{(\ref{#1})}

\newcommand{\sign}{\mathop{\rm sign}}

\newcommand{\tr}{\mathop{\rm Tr}}

\title{
\vspace{-9mm} \rightline{\small ITEP-LAT/2003-17} \vspace{-2mm}
\rightline{\small August, 2003}
$q\bar q$  and $2q2\bar q$  systems in terms of P-vortices
\thanks{Talk presented by M.I.P.
at Lattice 2003, Tsukuba}
}
\author{V.G. Bornyakov\address[IHEP]{Institute for High Energy Physics,
        Protvino, 142284, Russia}$^,$\address[ITEP]{Institute of Theoretical
and  Experimental Physics, B.~Cheremushkinskaya~25, Moscow, 117259, Russia},
A.V. Kovalenko\addressmark[ITEP], M.I. Polikarpov\addressmark[ITEP] and D.A.
Sigaev\addressmark[ITEP]\thanks{Work is partially supported by grants RFBR
02-02-17308, \uppercase{RFBR 01-02-17456, DFG-RFBR 436 RUS 113/739/0,
INTAS-00-00111} and \uppercase{CRDF} award \uppercase{RPI-2364-MO}-02.} .}
\begin{document}
\begin{abstract}
We study the action and the energy densities of the confining string in the
indirect $Z(2)$~projection of $SU(2)$ lattice gauge theory. We find that the
width of the
confining string is proportional to the logarithm of the distance between the
quark and antiquark. In $2q2\bar{q}$ system we observe the effect of the
reconstruction of the flux tube when we change the distance between two
$q\bar{q}$ pairs.
\end{abstract}

\maketitle
\section{INTRODUCTION}
The structure of the confining string in terms of the Abelian or monopole
operators after the  Abelian projection has been studied in lattice $SU(2)$
and $SU(3)$
gluodynamics and in lattice QCD with two quark
flavors~\cite{Bali:1997cp,Bornyakov:2001nd}.
The distribution of the monopole currents and the electric field near the
confining
string is similar to the distribution of the currents of the Cooper pairs and
magnetic field near the Abrikosov string~\cite{AnatomyKoma}. Thus the {\it
quantum} confining string is similar to the {\it classical} solution of the
equations of motion of the Abelian Higgs model. For the center--vortex model of
confinement~\cite{gfo} there is no corresponding classical model,
confinement is due to (quasi)randomly distributed magnetic fluxes in the
vacuum. On the other hand the numerical investigation of the confining string
in terms of P--vortices is even simpler than that in the Abelian projection,
since
the statistical fluctuations are smaller~\cite{Bornyakov:ej}. In Section~2 we
give the main definitions. In Section~3 we present the results of the
calculations of the electric and magnetic fields near the confining string. In
Section~4 we discuss the confining strings in the $2q2\bar{q}$ system.

All results are obtained  on $24^4$ lattice at
$\beta=2.40$ (20 statistically independent configurations), $\beta=2.50$ (50
configurations), and on $28^4$ lattice at $\beta=2.55$ (50 configurations),
$\beta=2.60$ (25 configurations). To fix the physical scale we use the
data for the string tension in lattice units~\cite{Fingberg:1992ju} and use
$\sqrt\sigma = 440\, MeV$.

\section{MAIN DEFINITIONS}
To extract P--vortices we use the indirect maximal center projection defined as
follows~\cite{gfo}. First, we fix the maximal Abelian gauge by maximizing
the functional $R_{U(1)}
=\sum_{x,\mu}\tr[U_{x\mu}\sigma^3U_{x\mu}^{+}\sigma^3]$ using the simulated
annealing algorithm~\cite{SA}. Then we replace link matrices $U_{x\mu}$ by the
Abelian link variables $U_{x\mu}\rightarrow
u_{x\mu}=\frac{U_{x\mu}^{11}}{|U_{x\mu}^{11}|}$. Further gauge fixing is
done by maximization of the functional $R_{Z(2)} =\sum_{x,\mu} [\Re e\;
u_{x\mu}]^2$. Finally, the maximal center projection is performed by replacing
$u_{x\mu}\rightarrow Z_{x\mu}=\sign\;\Re e\; u_{x\mu}$.

Since we are working on the symmetric lattices we have no apriori defined time
axis. But when we introduce the $R\times T$ Wilson loop in the $i,4$ plane we
define the time direction ($\mu=4$) and it becomes possible to define the
lattice electric and magnetic field strength as follows:

\beqn\label{E} (\vec{E}^a_{lat}(x))^2  =  \sum_{i=1}^{3} \left(1 -
P_{4i}(x)\right)\, ,\\
(\vec{H}^a_{lat}(x))^2  =  \sum_{i>j} \left(1 - P_{ij}(x)\right)\, ,\label{H}
\eeqn
where plaquette $P_{\mu\nu}(x)= \frac12 \mbox{Tr} U_{x,\mu\nu}$ in case of $SU(2)$ fields.
The Euclidean action density and energy density are defined as follows:

\beqn\label{sedef} \sigma(x)=(\vec{E}^a_{lat}(x))^2 +(\vec{H}^a_{lat}(x))^2\, ,\\
\varepsilon(x)=(\vec{E}^a_{lat}(x))^2-(\vec{H}^a_{lat}(x))^2\, .\nonumber
\eeqn
In the $Z(2)$ projection the static quarks are represented by $Z(2)$ Wilson
loop, $W_{Z(2)}$, constructed from $Z(2)$ links. $P_{\mu\nu}(x)$ entering
\eq{E}, \eq{H} are now the average of four $Z(2)$ plaquettes in the
$\{\mu,\nu\}$ plain which share the site $x$. To measure the
expectation value of the operator $O(x)$ in the presence of static quarks the
following quantity should be calculated:

\begin{equation}\label{OWdef}
<O(x)>_{W} = \frac{<O(x) W>}{<W>} - <O(x)>\, ,
\end{equation}
where $W$ is the Wilson loop.
In ref.\cite{Bornyakov:ej} this expression was used to calculate the expectation
values of $Z(2)$ plaquettes in the vicinity of the confining string.

\section{NUMERICAL RESULTS}
\begin{figure}
\begin{center}
\includegraphics[scale=0.32]{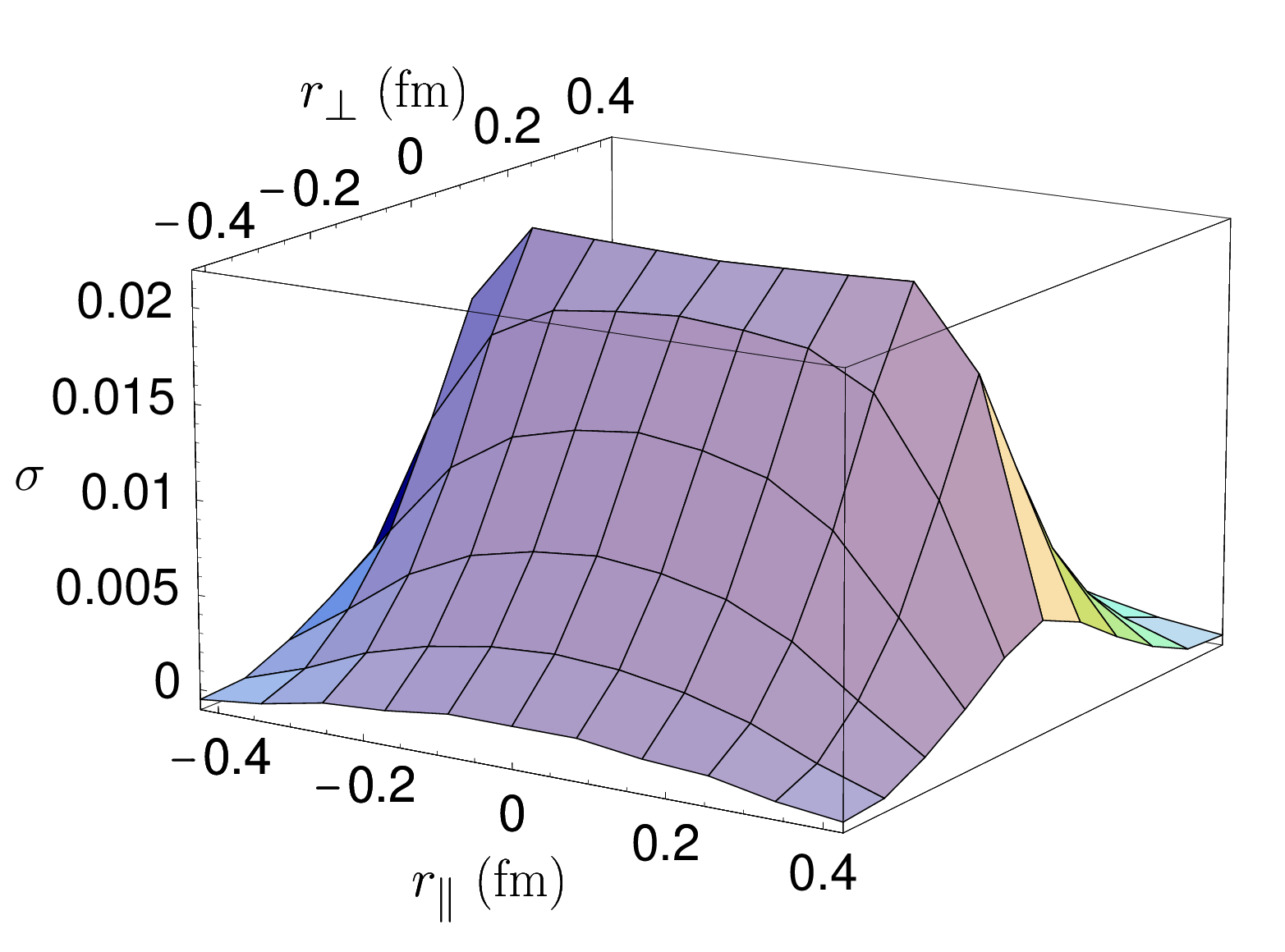}
(a)
\includegraphics[scale=0.32]{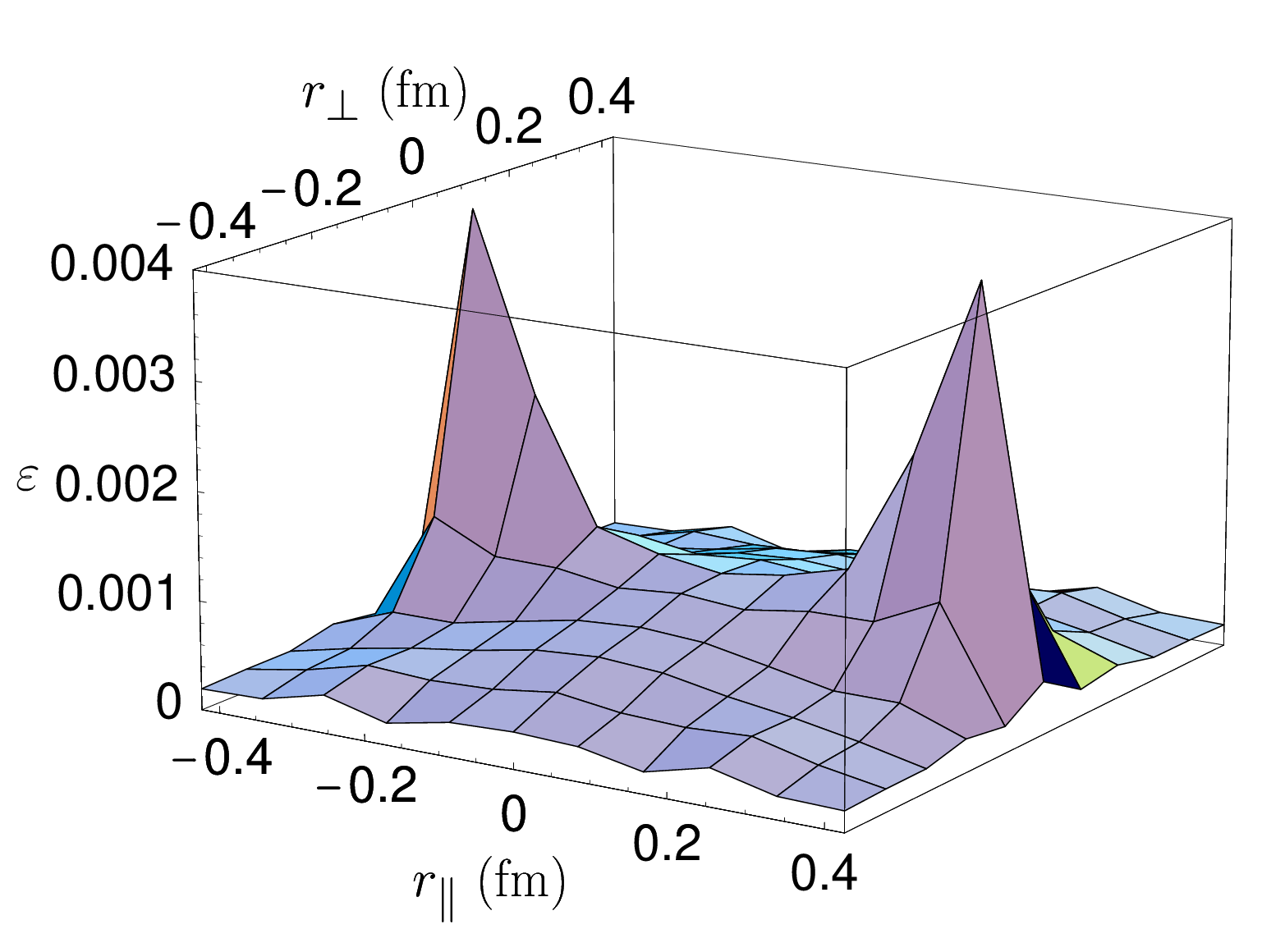}
(b) \end{center} \caption{$<\sigma(x)>_{W}$ (a) and $<\varepsilon (x)>_{W}$ (b)
defined by eqs.
\eq{sedef}, \eq{OWdef}. Calculations are performed
for $\beta=2.5$. The size of the Wilson loop is $8\times 10$.} \label{se3d}
\end{figure}

The results are presented in Figures~\ref{se3d} (a),(b).
The Coulomb peaks are not seen for the action density.
For the energy density one can see rather narrow peaks which is in
accordance with the result of ref.~\cite{gfo} where it was found that the
static potential in the $Z(2)$ projection is linear down
to very small distances.

\begin{figure}
\epsfxsize=0.43\textwidth \epsfbox{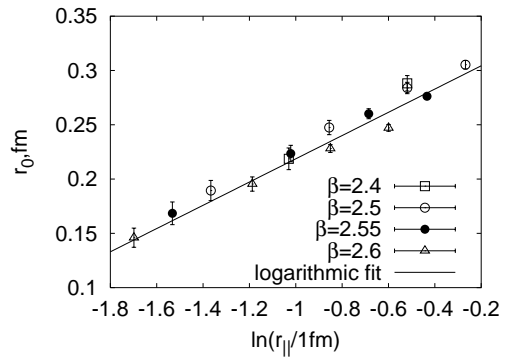} \caption{The confining string
radius as a function of $\ln r_\|$.} \label{r0}
\end{figure}

We define the radius of the $Z(2)$ confining string fitting the transverse
profile of the action density by the expression: $\sigma(r_\perp) = A
\exp\left[-\left(\frac {r_\perp}{r_0}\right)^2\right]$, where $A$ and $r_0$ are
the fitting parameters. We treat $r_0$ as the radius of the confining string.
To determine the dependence of the radius $r_0$ on the distance $r_\|$ between
the test quark and antiquark we used the data for four different $\beta$ and
various sizes of Wilson loops. Simulations at various $\beta$ yield results
which are in a good agreement (see Fig.~\ref{r0}), thus our results correspond
to the continuum limit.
To describe the dependence of $r_0$ on $r_\|$ we used the fitting function
$r_0(r_\|) = a\ln(r_\|/1\,{\rm fm})+b$. The results are: $a=0.11(1)$ fm,
$b=0.33(1)$ fm, $\chi^2/n_{dof}=4.31$.

\section{$2q2\bar q$  system}
\begin{figure}
\begin{center}
\includegraphics[scale=0.32]{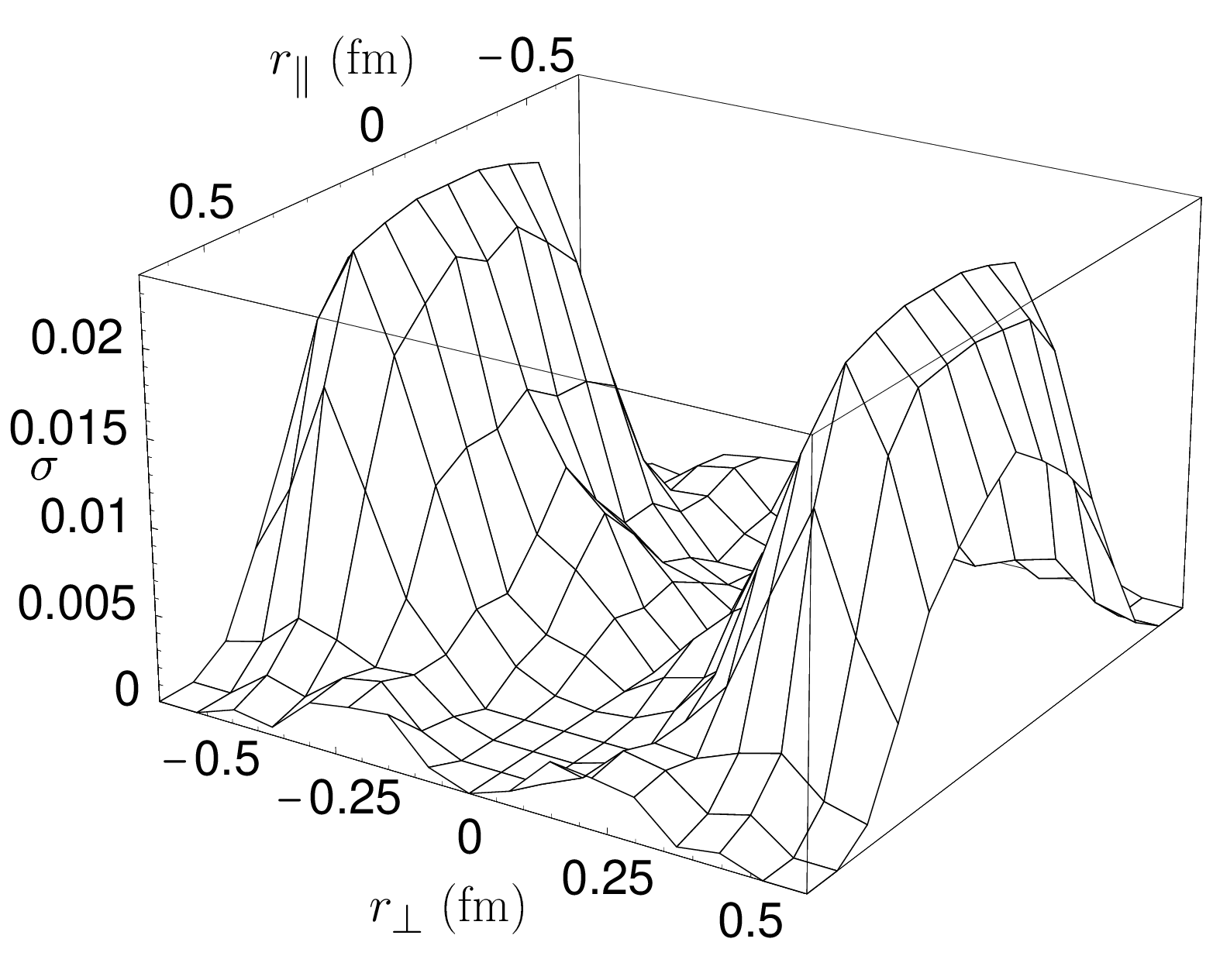}
(a)
\includegraphics[scale=0.32]{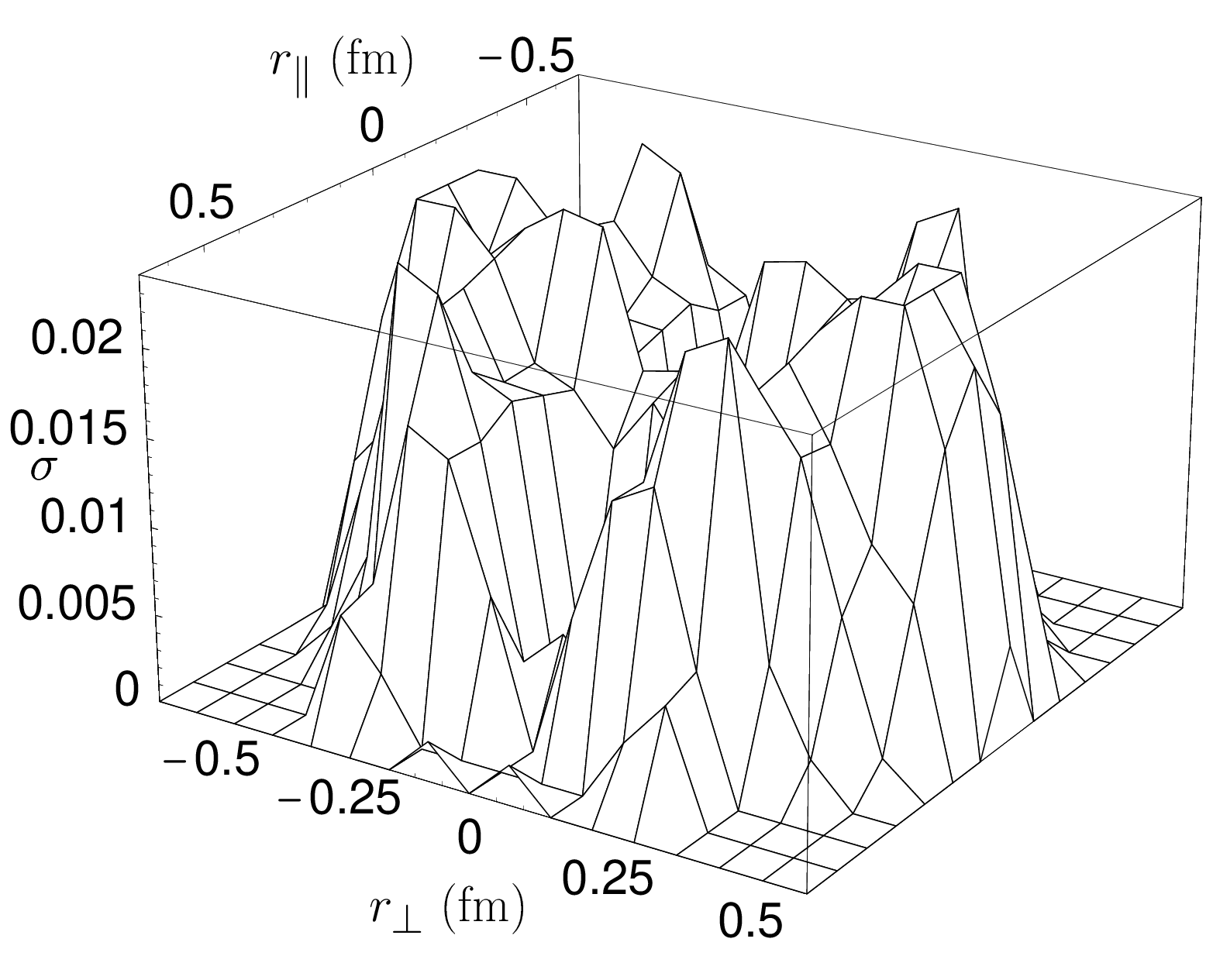}
  (b)
\includegraphics[scale=0.32]{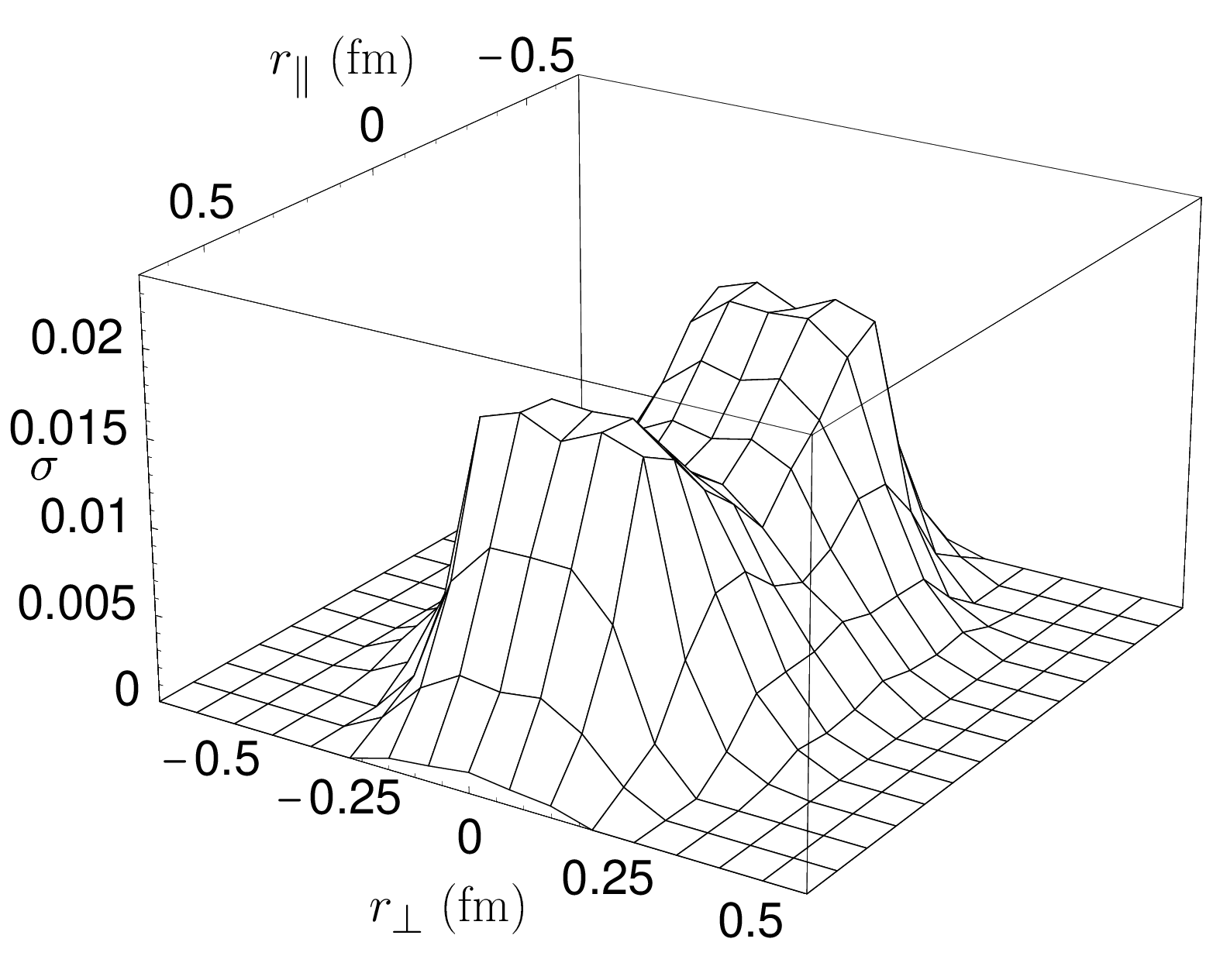}
(c)
\end{center} \caption{The $Z(2)$ action density for $2q2\bar q$ system at $\beta=2.5$.
The size of the Wilson loops is $8\times 10$.} \label{4q}
\end{figure}

The action and energy distribution for four static quarks in the $SU(2)$
gluodynamics has been studied in \cite{Pennanen:1998nu}. In Fig.\ref{4q} we
show the $Z(2)$ action density for two mesons formed by two static $q\bar q$
pairs. If the distance between quarks in a meson, $r_\|$, is smaller than the
separation of two mesons, $r_\bot$, one can see two flux tubes (see
Fig.\ref{4q}(a)). When $r_\| = r_\bot$ there is no any sign of the flux tubes
(see Fig.\ref{4q} (b)), the fluctuations are large. When $r_\bot$ becomes
smaller than $r_\|$ we observe that the flux tubes are formed in the direction
perpendicular to the initial one (see Fig.\ref{4q} (c)). Thus we observe the
effect of the reconstruction of the flux tubes.

\end{document}